\documentclass[final,5p,twocolumn,times,sort&compress]{elsarticle}

\def\preprintdate{IUHET 569, August 2012}

\usepackage{graphicx}

\def\al{\alpha}
\def\be{\beta}
\def\ga{\gamma}
\def\de{\delta}
\def\ep{\epsilon}

\def\et{\eta}

\def\ka{\kappa}
\def\la{\lambda}

\def\rh{\rho}

\def\si{\sigma}
\def\vs{\varsigma}

\def\up{\upsilon}
\def\ph{\phi}

\def\Ga{\Gamma}

\def\cR{{\cal R}}

\def\fr#1#2{{{#1} \over {#2}}}
\def\frac#1#2{{\textstyle{{#1}\over {#2}}}}
\def\half{{\textstyle{1\over 2}}}

\def\lsim{\mathrel{\rlap{\lower4pt\hbox{\hskip1pt$\sim$}}
    \raise1pt\hbox{$<$}}}
\def\gsim{\mathrel{\rlap{\lower4pt\hbox{\hskip1pt$\sim$}}
    \raise1pt\hbox{$>$}}}
\def\sqr#1#2{{\vcenter{\vbox{\hrule height.#2pt
         \hbox{\vrule width.#2pt height#1pt \kern#1pt
         \vrule width.#2pt}
         \hrule height.#2pt}}}}

\def\prt{\partial}
\def\pt#1{\phantom{#1}}

\def\etal{{\it et al.}}

\def\lrpartial{\raise 1pt\hbox{$\stackrel\leftrightarrow\partial$}}

\newcommand{\beq}{\begin{equation}}
\newcommand{\eeq}{\end{equation}}
\newcommand{\bea}{\begin{eqnarray}}
\newcommand{\eea}{\end{eqnarray}}
\newcommand{\rf}[1]{(\ref{#1})}
\newcommand{\nn}{\nonumber\\}

\def\rif{Riemann-Finsler}
\def\prif{pseudo-Riemann-Finsler}
\def\rie{Riemann}
\def\prie{pseudo-Riemann}
\def\fin{Finsler}
\def\pfin{pseudo-Finsler}
\def\ran{Randers}
\def\pran{pseudo-Randers}
\def\min{Minkowski}
\def\ber{Berwald}

\def\gram#1#2{{\rm gram}(#1,#2)}
\def\norm#1{\| #1 \|}

\def\tm#1{TM\backslash #1}

\def\Spe{S^\perp}
\def\Speo{S_1{}^\perp}

\def\m2{m^2}
\def\a2{a^2}
\def\b2{b^2}

\def\u2{u^2}

\def\up2{u^\perp^2{}}

\def\yy{y}
\def\y2{y^2}

\def\ypa{y^\parallel{}}
\def\ype{y^\perp{}}
\def\yp2{y^\perp{}^2}

\def\ssi{\rh}

\def\yj{{\yy^j}}
\def\yk{{\yy^k}}
\def\yl{{\yy^l}}
\def\ym{{\yy^m}}

\def\tch#1#2#3{\widetilde{\ga}_{#1 #2 #3}}
\def\hch#1#2#3{\widehat{\ga}_{#1 #2 #3}}

\def\tchu#1#2#3{\widetilde{\ga}^{#1}_{\pt{#1} #2 #3}}
\def\hchu#1#2#3{\widehat{\ga}^{#1}_{\pt{#1} #2 #3}}

\def\dt{\bullet}
\def\odt{\circ}

\def\cov{\widetilde D}

\def\S#1#2{s_{#1,#2}}
\def\R#1{r_{#1}} 
\def\Rpe#1{r_{#1}^\perp}
\def\A#1{a_{#1}} 
\def\B#1{b_{#1}} 
\def\H#1#2{H_{#1,#2}}
\def\Hpe#1#2{H_{#1,#2}^\perp}

\def\hperp{H^\perp{}}

\def\Fpe{{F^\perp}}

\def\Gpe{{G^\perp}}
\def\FHpe{{F_H^\perp}}

\def\sipe{\si^\perp{}}

\def\spe{s^\perp{}}

\def\gpe{g^\perp{}}
\def\Bf{S}

\def\Ipe{I^\perp{}}

\def\Mpe{M^\perp{}}

\def\kape{\ka^\perp{}}

\def\lam{\vs} 
\def\lamh{\et} 

\begin{document}

\begin{frontmatter}

\title{Bipartite \rif\ geometry and Lorentz violation}

\author{V.\ Alan Kosteleck\'y$^a$,
N.\ Russell$^b$,
and R.\ Tso$^c$}

\address{$^a$Physics Department, Indiana University,
Bloomington, IN 47405, U.S.A.\\
$^b$Physics Department, Northern Michigan University,
Marquette, MI 49855, U.S.A.\\
$^c$Physics Department, Embry-Riddle Aeronautical University,
Prescott, AZ 86301, U.S.A.
} 

\address{}
\address{\rm 
\preprintdate;
accepted for publication in Physics Letters B
}

\begin{abstract}

Bipartite Riemann-Finsler geometries
with complementary Finsler structures are constructed.
Calculable examples are presented
based on a bilinear-form coefficient for explicit Lorentz violation.

\end{abstract}

\end{frontmatter}

A famous example of \rif\ geometry is
\ran\ geometry \cite{gr},
which involves a \rie\ metric enhanced by a 1-form.
Its popularity stems partly from its simplicity
and calculability,
with relatively compact expressions attainable
for many geometric quantities
(see, e.g., Ref.\ \cite{bcs}).
It also has multiple links to physical situations.
Perhaps the simplest example involves
a relativistic charged massive particle
minimally coupled to a background electromagnetic 1-form potential
in (3+1)-dimensional spacetime,
for which the possible motions 
lie along the geodesics of a \pran\ metric.

A large class of \rif\ geometries,
including \ran\ geometry,
has recently been linked to Lorentz and CPT violation
in realistic effective field theory
\cite{ak-finsler}.
The basic idea is that motions of classical particles
in the general realistic effective field theory 
for Lorentz and CPT violation in curved spacetime,
the Standard-Model Extension (SME) 
\cite{akgrav},
follow geodesics in \prif\ spacetimes 
from which corresponding \rif\ geometries can be constructed. 
The Lorentz and CPT violation could arise 
in a fundamental theory unifying quantum physics and gravity 
such as strings 
\cite{ksp},
with the SME describing the resulting effects at attainable energies 
\cite{kp,owg}.
These notions about \rif\ geometries 
have application in a variety of related contexts 
\cite{dcpm-finsler,clp,gyb,rsv,sv,aknr,cmm,jsmv,nm,%
cjlw,ps,rgt,lph,kss}.

Among the novel geometries are \rif\ spaces 
of simplicity and calculability comparable to the \ran\ case.
One surprise is the existence of another calculable \rif\ space,
termed $b$ space,
which is also determined by a 1-form 
and has \fin\ structure complementary to that of \ran\ space.
Physically,
the corresponding \prif\ geometry 
is associated with the geodesic motion 
of a fermion in the presence of chiral CPT-odd Lorentz violation 
in (3+1)-dimensional \prie\ spacetime
\cite{aknr}.

In this note,
we explore the existence of other complementary pairs 
of \fin\ structures in this class of geometries.
For bipartite \fin\ structures 
constructed from the \rie\ metric $r_{jk}$
and a nonnegative symmetric bilinear form $s_{jk}$,
$j,k=1,2,3,\ldots,n$,
we show that when $s_{jk}$ has a single positive eigenvalue
the corresponding \rif\ geometry has a natural complement.
Some properties of these bipartite spaces are derived,
including the connection between 
$r$-parallel and \ber\ spaces.
As explicit examples,
we examine special cases of $H$ spaces
that have complementary bipartite structures of this type.
In (3+1)-dimensional spacetime,
the corresponding \prif\ structure governs
the geodesic motion of a fermion 
in the presence of CPT-even Lorentz violation.
We also identify isomorphisms between 
\ran\ space, $b$ space, $H$ space, and $\hperp$ space.
The notation and conventions adopted below 
are those of Ref.\ \cite{ak-finsler}.

A bipartite structure 
is a particular function on the tangent bundle $TM$
of the background spacetime manifold $M$.
In terms of $n$-dimensional positions $x^j$ and velocities $y^j$,
this function $F(x,y)$ takes the form
\cite{ak-finsler}
\beq
F(x,y) = \rh + \si,
\quad
\rh := \sqrt{y^j r_{jk} y^k} ,
\quad
\si := \pm \sqrt{y^j s_{jk} y^k} ,
\label{sstructure}
\eeq
where either sign of $\si$ can be chosen.
Both $r_{jk}$ and $s_{jk}$ are generically functions of $x^j$,
and indeed in the corresponding \prif\ geometries 
a position dependence of the SME coefficients
is natural in a gravitational background
\cite{akgrav,qbakpn,lvmm,cardinal,mds}.
Note that using the inverse \rie\ metric $r^{jk}$
to raise an index on $r_{jk}(x)$ and $s_{jk}(x)$ 
produces linear operators
$r^j{}_k (x) \equiv \de^j{}_k$ (the Kronecker delta)
and $s^j{}_k (x)$,
respectively.

The bipartite structure $F$ 
is positive for the positive sign of $\si$
and is positive for the negative sign of $\si$
when the nonzero eigenvalues of $s^j{}_k$ are less than one,
corresponding to the assumption of perturbative Lorentz violation.
Also, $F$ is positive homogeneous in $y^j$ of order one.
Moreover, 
$F$ is $C^\infty$ regular on the slit tangent bundle $\tm S$,
where $S=S_0 \cup S_1$ includes the usual slit 
$S_0 =\{ y: y^j = 0 \}$
and the slit extension 
$S_1 =\{ y: s^j{}_k y^k = 0, y^j\neq 0 \}$.
Typically,
$F$ is $y$ local,
but for certain choices of $s_{jk}$ 
the slit extension $S_1$ is empty
and $F(x,y)$ becomes $y$ global. 

With the above conditions, 
the bipartite structure $F$
becomes a \fin\ structure if it has strong convexity,
which occurs when 
the corresponding \fin\ metric $g_{jk}$
is positive definite on $\tm S$.
This metric is readily calculated to be
\beq
g_{jk} = \fr F \rh r_{jk} - \rh\si \ka_j \ka_k + \fr F \si s_{jk} ,
\label{metric}
\eeq
where
$\ka_j := {\rh_\yj}/\rh - {\si_\yj}/\si$.
We show below that for the cases of interest here
$g_{jk}$ is indeed positive definite on $\tm S$.
A more general result establishing conditions on $s_{jk}$
sufficient for strong convexity of $F$ would be of interest.

For the bipartite \fin\ structure,
the Cartan torsion is found to take the simple form
\beq
C_{jkl} =
- \half\rh\si \sum_{(jkl)} \ka_j\ka_{kl},
\label{cartan}
\eeq
where the sum spans cyclic permutations of $j$, $k$, $l$.
Here,
$\ka_{jk} := {\rh_{\yj\yk}}/{\rh} - {\si_{\yj\yk}}/{\si}$
involves the second $y^j$ derivatives of $\rh$ and $\si$.
Since the Cartan torsion is nonzero whenever $\si$ is nontrivial,
the Deicke theorem 
\cite{deicke}
implies the bipartite structure
is then noneuclidean as a \min\ norm.
With nonzero $\si$,
the bipartite geometry therefore cannot be a \rie\ geometry.

Our interest in this work lies in special bipartite geometries
that appear in complementary pairs.
To investigate this explicitly,
in what follows we restrict $s^j{}_k(x)$ to have rank $m$
with one nonzero positive eigenvalue $\lam (x)$
of multiplicity $m$, 
where $\lam < 1$ for $F$ to be positive on $\tm S$.
It follows that $s^j{}_k = \lam \hat s^j{}_k$,
where $\hat s^j{}_k$ is idempotent, 
$\hat s^2 = \hat s$.
Note that in an appropriate basis
$\hat s^j{}_k$ is the diagonal matrix with $m$ unit entries
and $n-m$ zero entries, 
$\hat s = I_m$. 
We thus have 
\beq
s^2 = \lam s,
\quad 
0< \lam <1.
\label{eigen}
\eeq
Note also that if $m=0$ then $s_{jk}=0$
and the geometry is Riemann,
while if $m=n$ then $s_{jk}=\lam r_{jk}$
and the geometry is again Riemann
but with a metric scaled by $(1\pm \sqrt{\lam})^2$.

To show strong convexity of $F$ for $s_{jk}$
satisfying the condition \rf{eigen},
which amounts to showing positive definiteness 
of the \fin\ metric \rf{metric} in this limit,
consider the determinant of $g_{jk}$. 
Some calculation reveals it can be written as
\beq
\det(g_{jk}) =
\left(\fr F \rh\right)^{n+1}
\left(\fr \Bf \si\right)^{m-1}
\det(r_{jk}) ,
\label{det}
\eeq
where the function $\Bf := \lam \rh + \si$
generalizes the function $B$ of Ref.\ \cite{ak-finsler}
and is always nonzero for $y^j\neq 0$,
with its sign matching the sign of $\si$.
The standard argument 
\cite{bcs}
for positive definiteness of $g_{jk}$ can then be applied.
With $F_{\ep} = \ssi + \ep \si$,
Eq.\ \rf{det} shows $g_\ep{}_{jk}$ has no vanishing eigenvalues
because $\det g_\ep >0$.
The eigenvalues of $g_\ep{}_{jk}$ are positive for $\ep=0$,
while no eigenvalue changes sign as $\ep$ grows to 1
because none vanishes.
This line of reasoning also confirms invertibility of $g_{jk}$.

The comparative elegance of the result \rf{det}
is reminiscent of the analogous expressions
for \ran\ space
\cite{bcs}
and $b$ space
\cite{ak-finsler}.
In fact,
\ran\ space is covered by two copies 
of the bipartite space \rf{sstructure} with opposite signs
and with $s_{jk} = a_j a_k$,
while the $b$ structure is a special case of Eq.\ \rf{sstructure}
with $s_{jk} = \b2 r_{jk} - b_j b_k$.
The result \rf{det} for these cases is related
via Theorem 2.3 of Ref.\ \cite{ss}
to the metric determinant for general $(\al,\be)$ spaces,
which have \fin\ structures of the form 
$F_{(\al,\be)}= \al \ph(\be/\al)$ 
for some $C^\infty$ positive function $\ph$,
where $\al =\rh$ and $\be$ is a 1-form on $\tm S$.
In this context,
the \ran\ structure $F_a$ appears as an $(\al,\be)$ structure
with $\al =\rh$, $\be = a\cdot y$, and $\ph = 1 + \be/\al$.
Also,
Shen has observed 
\cite{shenpc}
that the $b$ structure $F_b$ 
with constant norm $\norm{b}$
can be viewed as an $(\al,\be)$ structure
with $\al =\rh$, $\be = b\cdot y/\norm{b}$,
and $\ph = 1 \pm \norm{b} \sqrt{1 - (\be/\al)^2}$,
with metric determinant given 
by Lemma 1.1.2 of Ref.\ \cite{sczs}.
Even for the more complicated $F_{ab}$ structure of $ab$ space
\cite{ak-finsler},
a relatively compact result exists for the metric determinant.
Javaloyes and S\'anchez have recently studied
more general homogeneous functionals 
of \fin\ structures and 1-forms
\cite{js},
including the $(F_0,\be)$ spaces generated as
$\be$-deformations of a \fin\ structure $F_0$
\cite{shibata}
and the $(F_1,F_2)$ spaces 
generated by combining two \fin\ structures $F_1$ and $F_2$.
The $F_{ab}$ structure 
is a special case of an $(F_0,\be)$ structure
with $F_0=F_b$, $\be = a\cdot y$,
and $\ph = 1 + \be/F_0$,
so the metric determinant is given by
Proposition 4.24 of Ref.\ \cite{js}.
Modulo possible technical issues with the slit extension $S_1$,
the bipartite structure discussed here
takes the form of an $(F_1,F_2)$ space 
with $F_1=\rh$, $F_2 = \si$, and $\ph = 1 + F_2/F_1$,
although the result \rf{det} appears unexpectedly simple
given that $F_2$ is constructed from a bilinear form $s_{jk}$.
Together with the existence of numerous other \fin\ spaces
arising from the motion of fermions in the SME
\cite{ak-finsler},
this simplicity suggests 
that further attractive \fin\ geometries 
related to Lorentz violation in effective field theory
remain to be discovered.

If the rank $m$ of $s$ is nonextremal,
$0<m<n$,
then the image and kernel subspaces of $s$ are nontrivial.
Since these spaces are orthogonal,
we can uniquely project any vector $y$ in $TM_x$ 
into components parallel and perpendicular to the image subspace,
\beq
\ypa := \frac 1 {\lam} s y ,
\quad
\ype := y-\ypa.
\label{proj}
\eeq
Since $\ypa r \ype=0$,
the three vectors $y$, $\ypa$, and $\ype$
can be viewed as forming a right-angle triangle.
Their norms satisfy the inequalities 
$\norm \ypa \leq \norm y$ and $\norm \ype \leq \norm y$,
which are useful for several purposes.
For example,
for $s$ obeying Eq.\ \rf{eigen} with $\lam<1$,
the bipartite structure $F$ is positive.
This result can be viewed as a consequence of the inequality
$\norm y > \norm \ypa$ for $y\neq \ypa$,
which implies
$\rh > \sqrt{ysy}/{\sqrt{\lam}}$,
hence $\rh - \sqrt{ysy} > 0$,
and thus $\rh + \si > 0$.
As another example,
we can apply the inequality $\norm \ypa \leq \norm y$ to show 
the sign of the function $\Bf$ introduced in Eq.\ \rf{det}
matches that of $\si$,
a result used above to prove strong convexity of $F$.
For positive $\si$,
$\Bf$ is positive by inspection.
Noting that $\si = \pm \sqrt{\lam}\norm{\ypa}$,
for negative $\si$
we can write $\Bf = \lam \norm{y} - \sqrt{\lam} \norm{\ypa}
\leq \sqrt{\lam}(\sqrt{\lam} -1) \norm{y} <0$.
The sign of $S/\si$ is therefore always positive,
as claimed.

In terms of the projected vectors \rf{proj},
the contribution $\si$ to the bipartite structure $F$
can be written in the form 
$\si = \pm \sqrt{\lam}\sqrt{y r \ypa}$.
However,
the vectors $\ypa$ and $\ype$ play analogous roles 
in the triangle. 
This suggests the perpendicular component $\ype$
can be used to define a complementary bipartite structure $\Fpe$
given by
\beq
\Fpe := \rh + \sipe,
\quad
\sipe := \pm \sqrt{\lam}\sqrt{y r\ype} = \pm \sqrt{\lam \y2 - ysy},
\label{Fperp}
\eeq
where the sign choice can be independent of that adopted for $\si$.
Up to a possible sign,
the map $F\to \Fpe$ is thus implemented by
the replacement
\beq
s\to \spe := \lam r - s,
\label{map}
\eeq
which induces $\si\to\sipe$, 
$\Bf\to\Bf^\perp= \lam \rh + \sipe$,
$\ka_j\to\kape_j = {\rh_\yj}/\rh - {\sipe_\yj}/\sipe$,
etc.
For example,
using this replacement 
the corresponding \fin\ metric $\gpe_{jk}$,
its determinant $\det(\gpe_{jk})$,
and the Cartan torsion $C^\perp{}_{jkl}$ 
can be obtained from Eqs.\ \rf{metric}, \rf{det},
and \rf{cartan},
respectively. 
Note that a second iteration 
recovers $s$, $s\to \lam r - s \to s$,
so the replacement \rf{map} is a reflection.
Also, in terms of the idempotent linear operator $\hat s^j{}_k$ 
the replacement gives $\hat s \to I - \hat s$,
so in a suitable basis it amounts to the substitution 
$I_m\to I_{n-m}$. 

With $0<\lam<1$ as before,
the inequality $\norm y > \norm\ype$ for $y\neq\ype$
implies that $\Fpe$ is positive on $\tm S$.
Also, 
$\Fpe$ is positive homogeneous in $y^j$ of order one,
and it is $C^\infty$ regular on the slit tangent bundle $\tm \Spe$,
where $\Spe=S_0 \cup \Speo$ 
involves the perpendicular slit extension 
$\Speo =\{y: s^j{}_k y^k = \lam y^j, y^j \neq 0 \}$.
Moreover,
applying the standard argument
\cite{bcs}
to the determinant $\det(\gpe_{jk})$ 
verifies that $\Fpe$ has strong convexity.
These results imply that $\Fpe$ is a Finsler structure.

The above line of reasoning shows 
that bipartite Finsler structures 
obeying the condition \rf{eigen} always appear 
in complementary pairs, 
$F$ and $\Fpe$.
One example of such a pairing
is provided by the \ran\ structure $F_a$
and the $b$ structure $F_b$
\cite{ak-finsler}.
Another example involving $H$ space is presented below.

We remark in passing that both $F$ and $\Fpe$ 
can be expressed in terms of the Gram determinant or gramian,
which for two vectors $y$, $z$ is 
$\gram y z = y^2 z^2 - (y\cdot z)^2$.
Noting that $\gram y {sy} = \si^2 \sipe^2$,
we find
\beq
\hskip -12pt
F= \rh \pm \sqrt{\gram y {sy/\sipe}} ,
\quad
\Fpe= \rh \pm \sqrt{\gram y {sy/\si}} .
\label{Sgram}
\quad
\eeq
This generalizes the gramian expressions
for $F_a$ and $F_b$ given in Ref.\ \cite{ak-finsler}. 

Using the determinant \rf{det},
we can calculate
the mean Cartan torsion $I_j = (\ln(\det g))_{y^j}/2$ for $F$, 
\beq
I_j = 
- \half \left[(n+1) \fr\si F 
- (m-1) \fr{\lam \rh}\Bf\right]\ka_j .
\label{mct}
\eeq
Combining this with the Cartan torsion \rf{cartan}
yields the Matsumoto torsion 
\bea
M_{jkl} =
- \half F \sum_{(jkl)} 
\ka_j \Big[\fr{m-1}{n+1}
\fr{\lam \rh}\Bf (\rh_{\yk\yl} + \si_{\yk\yl})
- \si_{\yk\yl}\Big] .
\label{mt}
\eea
The corresponding expressions $\Ipe_j$ and $\Mpe_{jkl}$
for the complementary bipartite structure $\Fpe$
can be obtained via the map \rf{map}.
They take the same forms \rf{mct} and \rf{mt}
with the substitutions
$F\to\Fpe$, 
$\si\to\sipe$, 
$\Bf\to\Bf^\perp$,
$\ka_j\to\kape_j$,
and $m\to n-m$. 

Except for special examples,
notably the rank-1 cases,
the Matsumoto torsions 
$M_{jkl}$ and $\Mpe_{jkl}$ are nonzero
and so the Matsumoto-H\=oj\=o theorem
\cite{mh}
shows that $F$ and $\Fpe$
typically differ from the \ran\ structure $F_a$
despite their apparent simplicity.
Moreover,
as we show explicitly below using $H$ space,
only a subset of the bipartite $F$ and $\Fpe$ structures
generate $b$ space.
Interesting novel cases are therefore 
contained within \fin\ structures 
built from $s_{jk}$ satisfying the condition \rf{eigen}. 
One intriguing open question in this context
is identifying a new torsion
that distinguishes $b$ space from other \fin\ spaces,
in analogy with the role of the Matsumoto torsion
in distinguishing \ran\ space from other \fin\ spaces. 
The simplicity of $b$ space,
the complementary nature of $F_b$ to the \ran\ structure $F_a$,
and the chirality relationship arising in the SME context
between the pseudo-\fin\ structures associated with $F_a$ and $F_b$
all are suggestive indications that such a torsion exists.

Since any $r$-parallel $b$ space is known to be \ber\ 
\cite{ak-finsler},
it is natural to ask whether
a similar result holds for $r$-parallel bipartite spaces
satisfying the condition \rf{eigen}.
We can investigate this and obtain some related results
by considering the geodesics associated with $F$,
which obey
\beq
F \fr d {d\la} \left(\fr 1 F \fr{d x^j}{d\la}\right) + G^j = 0 ,
\label{geodesic}
\eeq
where the spray coefficients 
$G^j := g^{jm} \Ga_{mkl} y^k y^l$
are defined in terms of the Christoffel symbol
$\Ga_{jkl}$ for $g_{jk}$.
The first step towards obtaining the spray coefficients $G^j$
is to evaluate $G_j$ using
$G_j = \Ga_{jkl} y^k y^l$.
We find 
\bea
G_j = \rh F \tch j \dt \dt
+ \rh^2 (\prt_\dt \si - \si \tch \dt\dt\dt ) \ka_j
+ \fr {\rh^2 F}{\si} \hch j \dt\dt ,
\eea
where a lower index $m$ contracted with $r^{mk}\rh_\yk$
is denoted by a bullet $\dt$,
with contractions external to any derivatives that appear.
The Christoffel symbol 
for the \rie\ metric $r_{jk}$ is denoted $\tch j k l$,
while that for $s_{jk}$ is denoted $\hch j k l$.
Note that some expressions involving $\hch j k l$
can be more compactly expressed
using the $r$-covariant derivative $\cov_j$ 
and the relationship
\beq
\hch j k l |_{\prt \rightarrow \cov_{}} := 
\half ( \cov_k s_{jl} + \cov_l s_{jk} - \cov_j s_{kl}) 
= \hch j k l - s_{jm} \tchu m k l .
\eeq

To find the spray coefficients $G^j$,
we need the inverse bipartite metric $g^{jk}$.
Since $g_{jk}$ is positive definite, 
the inverse metric exists.
After some calculation,
we find 
\beq
g^{kl} =
\fr \rh F \left(r^{kl} 
+ \fr{\sipe^2 \rh}{\si^2 \Bf} \la^k \la^l
- \fr \rh \Bf s^{kl} \right) ,
\label{inverse}
\eeq
where
\beq
\la_j := \fr 1 \sipe \left(s_{jk}y^k - \fr {\si \Bf} F \rh_j \right) .
\eeq
For the complementary structure $\Fpe$,
the inverse metric $\gpe_{jk}$
is again obtained via the replacement \rf{map}.
These results are similar in form 
to the expressions (22) and (23) of Ref.\ \cite{ak-finsler}
for the inverse \fin\ metric of $b$ space.

Using Eq.\ \rf{inverse},
a calculation shows that 
the bipartite spray coefficient $G^j$ can be written as 
\bea
\hskip -50 pt
&
G^j = \rh^2 \tchu j\dt\dt
+ \fr{\rh^3}{\Bf \si^3}
\big[
\si^3 \hchu j\dt\dt
+ \rh\si^2 \spe^{jk} \hch k\dt\dt
\nn
\hskip -50 pt
&
\qquad \qquad \qquad
\qquad \qquad \qquad
-\rh \sipe (\sipe\hch \odt\dt\dt 
+ \si \hch \dt\dt\dt)\la^j)
\big]_{\prt \rightarrow \cov_{}} ,
\quad
\label{spray}
\eea
where an index $\circ$ represents a lower index $m$ contracted with
$(s^\perp y)^m/\sipe$ externally to any derivatives.
Note that the replacement \rf{map}
can be used to obtain the expression 
for the complementary spray coefficients $\Gpe^j$,
which satisfy a geodesic equation for $\Fpe$
taking the form \rf{geodesic}.

The result \rf{spray} reveals that $G^j$ 
contains the standard term $\tch jkl y^j y^k$
together with a linear combination of terms,
each of which involves 
the Riemann covariant derivative acting on $s_{jk}$.
It follows from \rf{spray} that
if the bipartite form $s_{jk}$ is $r$-parallel, 
$\cov_l s_{jk}=0$,
then the spray coefficients $G^j$ reduce to the usual \rie\ case
and the trajectories satisfy the usual \rie\ geodesic equation.
In this situation the spray coefficients are quadratic in $y^j$,
so the third $y^j$ derivative of $G^j$ is zero,
and therefore the \ber\ h-v curvature
$^B{}P_k{}^j{}_{lm} := - F (G^j)_{\yk\yl\ym}/2$
vanishes.
We can conclude that
any $r$-parallel bipartite space satisfying the condition \rf{eigen}
is necessarily \ber.
The same result follows for the bipartite space 
with complementary structure $\Fpe$. 
It would be of interest to investigate 
the validity of the converse hypothesis 
that any bipartite \ber\ space obeying 
the condition \rf{eigen} is $r$-parallel.
In any case,
the result established above is consistent with the conjecture 
that any SME-based \rif\ space is \ber\ 
iff it has $r$-parallel coefficients for Lorentz violation
\cite{ak-finsler}.
Since the presence of nonzero $r$-parallel $s^j{}_k$
leaves geodesics unaffected,
the result also is indicative 
of the existence of a variable transformation or redefinition
that would eliminate $s^j{}_k$ in this limit,
just as certain unphysical coefficients can be eliminated
in the SME
\cite{akgrav,dcak,aknr,rl-redef,redef,km-nonmin}.
Investigation of these two open conjectures is likely to lead
to additional mathematical and physical insights.

\begin{table*}
\begin{tabular}{c||c@{\hspace{5pt}}cccccc@{\hspace{5pt}}c@{\hspace{5pt}}ccc}
$		$&$	$&$		$&$		$&$		$&$		$&$	m	$&$		$&$		$&$		$&$		$&$		$	\\
$	n	$&$	0	$&$	1	$&$	2	$&$	3	$&$	4	$&$	5	$&$	6	$&$	\ldots	$&$	n-2	$&$	n-1	$&$	n	$	\\
\hline																									
$	1	$&$	\R 1	$&$	\Rpe 1	$&$		$&$		$&$		$&$		$&$		$&$		$&$		$&$		$&$		$	\\
$	2	$&$	\R 2	$&$	\A 2 = \B 2	$&$	\Rpe 2	$&$		$&$		$&$		$&$		$&$		$&$		$&$		$&$		$	\\
$	3	$&$	\R 3	$&$	\A 3 = \Hpe 3 2	$&$	\B 3 = \H 3 2	$&$	\Rpe 3	$&$		$&$		$&$		$&$		$&$		$&$		$&$		$	\\
$	4	$&$	\R 4	$&$	\A 4	$&$	\H 4 2 = \Hpe 4 2	$&$	\B 4	$&$	\Rpe 4	$&$		$&$		$&$		$&$		$&$		$&$		$	\\
$	5	$&$	\R 5	$&$	\A 5 = \Hpe 5 4	$&$	\H 5 2	$&$	\Hpe 5 2	$&$	\B 5 = \H 5 4	$&$	\Rpe 5	$&$		$&$		$&$		$&$		$&$		$	\\
$	6	$&$	\R 6	$&$	\A 6	$&$	\H 6 2 = \Hpe 6 4	$&$	\S 6 3	$&$	\H 6 4 = \Hpe 6 2	$&$	\B 6	$&$	\Rpe 6	$&$		$&$		$&$		$&$		$	\\
$	\vdots	$&$		$&$		$&$		$&$		$&$		$&$		$&$		$&$	\vdots	$&$		$&$		$&$		$	\\
$	\mbox{odd}	$&$	\R n	$&$	\A n = \Hpe n {n-1}	$&$	\H n 2	$&$	\Hpe n {n-3}	$&$	\H n 4	$&$	\Hpe n {n-5}	$&$	\H n 6	$&$	\cdots	$&$	\Hpe n 2	$&$	\B n = \H n {n-1}	$&$	\Rpe n	$	\\
$	\mbox{even}	$&$	\R n	$&$	\A n	$&$	\H n 2 = \Hpe n {n-2}	$&$	\S n 3	$&$	\H n 4 = \Hpe n {n-4}	$&$	\S n 5	$&$	\H n  6 = \Hpe n {n-6}	$&$	\cdots	$&$	\H n {n-2} = \Hpe n 2	$&$	\B n	$&$	\Rpe n	$	\\
\end{tabular}
\caption{\label{mappings}
Isomorphisms between \rie, \ran, 
$b$, $H$, $H^\perp$, and bipartite spaces.}
\end{table*}

The $y$-derivative $p_j:= F_{y^j}$ of a \fin\ structure
plays an important role in both mathematics and physics.
Mathematically,
$p_j$ defines the Hilbert form $F_{y^j} dx^j$.
Physically,
the corresponding quantity for a \pfin\ structure
is the canonical momentum.
The $y$-derivative $p_j$ 
determines an algebraic variety $\cR(p)$,
which is the dispersion relation governing the geodesic motion.
For the bipartite structure \rf{sstructure},
$p_j$ takes the form 
$p_j = r_{jk}y^k/{\rh} + s_{jk}y^k/{\si}$.
Restricting attention to $F$ obeying the condition \rf{eigen},
we find the dispersion relation can be written as 
\beq
(p^2 -1 + \lam)^2 - 4 psp = 0.
\label{disprel}
\eeq
The corresponding result for the complementary structure $\Fpe$
is obtained by the replacement \rf{map}. 
For example, 
the dispersion relation for the \ran\ structure $F_a$ 
is given by Eq.\ \rf{disprel} with $s_{jk} = a_j a_k$,
while that for the $b$ structure $F_b$
follows when $s_{jk} = b^2 r_{jk} - b_j b_k$.
These expressions are the \fin\ versions
of the \pfin\ dispersion relations 
derived for the motion of a classical fermion
in the presence of nonzero SME coefficients $a_\mu$ and $b_\mu$ 
in (3+1)-dimensional spacetime
\cite{dcak},
the effects of which have been sought in numerous experiments
\cite{tables}.
General descriptions of Lorentz-violating dispersion relations
can be found in Refs.\ \cite{rl-fermion,badc,gls,km-nonmin}.

Another interesting SME coefficient is the 2-form $H_{\mu\nu}$,
which arises naturally in some models 
with spontaneous Lorentz breaking
\cite{abk}
and for which the dispersion relation is also known 
\cite{akrl}.
Physical effects from $H_{\mu\nu}$ have been studied 
in the electron sector using a torsion pendulum
\cite{electron},
in the neutron sector with a He-Xe comagnetometer
\cite{neutron},
in the muon sector in a storage ring
\cite{muon},
and in the neutrino sector using neutrino oscillations
\cite{neutrino}.
For the generic case
the form of the associated \pfin\ structure is presently unknown,
but the special case with vanishing quadratic invariant 
$Y =\ep^{\al\be\ga\de}H_{\al\be}H_{\ga\de}/8$
yields a calculable example
\cite{aknr}.
Associated with these \prif\ spaces is a \fin\ geometry,
$H$ space, that involves a 2-form $H_{jk}$ 
\cite{ak-finsler}.

Here,
we consider a bipartite limit of the $H$ geometry 
obtained via a suitable constraint on 
the linear operator $H^j{}_k = r^{jl}H_{lk}$.
The antisymmetry of $H_{jk}$ implies $H^j{}_k = - H_k{}^j$,
so $H^j{}_k$ has even rank.
In odd dimensions it therefore has at least one zero eigenvalue,
while the total number of zero eigenvalues 
is odd in odd dimensions and is even in even dimensions.
The quadratic product $(H^2)^j{}_k = H^j{}_l H^l{}_k$
obeys $(H^2)^j{}_k = (H^2)_k{}^j$,
and all its nonzero eigenvalues are negative.
We focus attention on the restricted class of $H^j{}_k$ 
for which $(H^2)^j{}_k$ 
has only a single nonzero eigenvalue $-\lamh$,
so that 
\beq
H^4 = -\lamh H^2 .
\label{Hcondition}
\eeq
Since the condition \rf{Hcondition}
is of the form \rf{eigen},
we may define a bipartite $H$ space
by identifying $s = -H^2$, $\lam = \lamh$.
The associated \fin\ structure $F_H$
and its complementary structure $\FHpe$ are
\beq
F_H=\rh \pm\sqrt{-yH^2y},
\quad
\FHpe=\rh \pm\sqrt{\lamh y^2 +yH^2y} ,
\eeq
where the sign choices in the two expressions can be independent.
In terms of the gramian,
we can write 
\beq
\hskip -10pt
F_H = \rh \pm \sqrt{\gram y{-H^2 y/\sipe}} 
= \rh \pm \sqrt{\gram y{Hy/\rh} } ,
\eeq
where the first expression is of the type \rf{Sgram} 
and the second exploits the antisymmetric nature of $H_{jk}$.

The basic properties of this restricted $H$ space
follow by applying the results 
for $s_{jk}$ satisfying the condition \rf{eigen}.
The \fin\ metric for $F_H$ takes the form \rf{metric}
with $s_{jk}= - (H^2)_{jk}$,
while the metric determinant is given by Eq.\ \rf{det}
and its inverse by Eq.\ \rf{inverse}.
The Cartan torsion and its mean,
the Matsumoto torsion,
the spray coefficients,
and the dispersion relation
are all given by substitution into formulae presented above. 
The analogous results for the complementary structure $\FHpe$
can be found by the replacement 
$(H^2)_{jk} \to \lamh r_{jk} + (H^2)_{jk}$.
Note that one key difference between the restricted $H$ space
and the bipartite space obeying the condition \rf{eigen}
is that the rank $m$ is necessarily even for $H$ space.
Note also that the complementary bipartite structure $\FHpe$
is the $n$-dimensional \fin\ analogue 
of the (3+1)-dimensional \pfin\ structure
given in Eq.\ (15) of Ref.\ \cite{aknr},
while the dispersion relation \rf{disprel} for $\FHpe$ 
is the $n$-dimensional \fin\ analogue 
of the (3+1)-dimensional dispersion relation for $Y=0$. 

As seen above,
\rie\ space,
\ran\ space,
$b$ space,
and the two restricted $H$ spaces
are all examples of bipartite spaces 
obeying the condition \rf{eigen}.
Any such bipartite space is fixed by specifying
the dimension $n$ of the configuration space,
the rank $m$ of $s^j{}_k$,
and the eigenvalue $\lam$.
This implies certain spaces are isomorphic.
For example, 
\ran\ space and $b$ space are isomorphic in two dimensions 
when $b_j=a_j$
because both have $n=2$, $m=1$, and $\lam = a^2$.
To express these isomorphisms compactly,
it is convenient to introduce notation for the various spaces.
For dimension $n$ and rank $m$,
let $\S n m$ be the bipartite space 
obeying the condition \rf{eigen}.
If $m=0$,
then it suffices to indicate $n$
and the space is \rie, 
denoted $\R n$.
The case $m=n$
yields the complementary \rie\ space with scaled metric,
written $\Rpe n$.
The rank $m$ is always 1 for the \ran\ spaces $\A n$,
while the rank $n-1$ is fixed by the dimension
for the $b$ spaces $\B n$.
The restricted $H$ space in $n$ dimensions 
with $(H^2)^j{}_k$ of rank $m$ is denoted $\H n m$,
and the complementary space is written $\Hpe n m$.

Using these conventions
and assuming a definite value of $\lam$,
Table \ref{mappings} summarizes
the isomorphisms between the various cases.
Each cell in the table represents an $\S n m$ space
with specified $n$ and $m$.
Note that cells with $m>n$ are meaningless and are left blank.
Most of the $\S n m$ spaces can be identified
with one or more of the other spaces,
so we use $\S n m$ only where no other notation applies.
Only for even $n$ with certain odd $m$ 
do $\S n m$ spaces exist that are distinct 
from the $\A n$, $\B n$, $\H n m$, and $\Hpe n m$ spaces.
The first three occurrences of this 
are $\S 6 3$ in six dimensions
and $\S 8 3$ and $\S 8 5$ in eight dimensions.
For ranks $m=0$ and $m=n$,
\rie\ spaces are obtained,
and these have no isomorphisms with other bipartite spaces 
because the Cartan torsion \rf{cartan} vanishes.
The rank-one \ran\ spaces $\A n$ 
in odd dimensions 
are isomorphic to the complementary $H$ spaces $\Hpe n {n-1}$,
while in even dimensions they are unique
except for the isomorphism with $b$ space for $n=2$.
Analogously,
the rank-$(n-1)$ spaces $\B n$
in odd dimensions are isomorphic to $\H n {n-1}$,
while in even dimensions they are unique except for $\B 2 = \A 2$.
For other ranks,
the $\S n m$ spaces in odd dimensions
generate an alternating series of restricted $H$ spaces
and their complements.
Also,
each restricted $H$ space with even rank and dimension
is isomorphic to a complementary $H$ space,
$\H n m = \Hpe n {n-m}$.
The general cases for odd and even dimensions 
are listed in the last two rows of the table.

As a final remark,
we note that the comparatively simple \fin\ structure
associated with bipartite geometries obeying the condition \rf{eigen}
and the variety of isomorphisms displayed 
in Table \ref{mappings}
together suggest the potential 
for interesting physical applications of Eq.\ \rf{sstructure}
in addition to the \prif\ applications to the SME mentioned above.
For example,
Shen 
\cite{zs,dbcr}
has demonstrated that \ran\ geodesics
correspond to solutions of the Zermelo navigation problem
of navigation control in an external wind
related to the coefficient $a_j$.
This result provides a direct physical application
of the spaces with $m=1$ listed in the third column 
of Table \ref{mappings}.
Finding analogous physical interpretations 
for the other entries in the table
is an intriguing open challenge.

\section*{Acknowledgments}

This work was supported in part
by the Department of Education 
under the McNair Scholars program,
by the Department of Energy
under grant number DE-FG02-91ER40661,
by the National Science Foundation
under the REU program,
and by the Indiana University Center for Spacetime Symmetries
under an IUCRG grant.


\begin{thebibliography}{xx}

\bibitem{gr}
G.\ Randers,
Phys.\ Rev.\ {\bf 59}, 195 (1941).

\bibitem{bcs}
D.\ Bao, S.-S.\ Chern, and Z.\ Shen,
{\it An Introduction to Riemann-Finsler Geometry},
Springer, New York, 2000.

\bibitem{ak-finsler}
V.A.\ Kosteleck\'y,
Phys.\ Lett.\ B {\bf 701}, 137 (2011).

\bibitem{akgrav}
V.A.\ Kosteleck\'y,
Phys.\ Rev.\ D {\bf 69}, 105009 (2004).

\bibitem{ksp}
V.A.\ Kosteleck\'y and S.\ Samuel,
Phys.\ Rev.\ D {\bf 39}, 683 (1989);
V.A.\ Kosteleck\'y and R.\ Potting,
Nucl.\ Phys.\ B {\bf 359}, 545 (1991).

\bibitem{kp}
V.A.\ Kosteleck\'y and R.\ Potting,
Phys.\ Rev.\ D {\bf 51}, 3923 (1995).

\bibitem{owg}
O.W.\ Greenberg,
Phys.\ Rev.\ Lett.\ {\bf 89}, 231602 (2002).

\bibitem{dcpm-finsler}
D.\ Colladay and P.\ McDonald, 
Phys.\ Rev.\ D {\bf 85}, 044042 (2012).

\bibitem{clp}
M.\ Cambiaso, R.\ Lehnert, and R.\ Potting,
Phys.\ Rev.\ D {\bf 85}, 085023 (2012).

\bibitem{gyb}
G.Yu.\ Bogoslovsky,
Int.\ J.\ Geom.\ Meth.\ Mod.\ Phys.\ {\bf 9}, 125007 (2012).

\bibitem{rsv}
J.M.\ Romero, O.\ Sanchez-Santos, and J.D.\ Vergara,
Phys.\ Lett.\ A {\bf 375}, 3817 (2011).

\bibitem{sv}
S.I.\ Vacaru,
Class.\ Quant.\ Grav.\ {\bf 28}, 215001 (2011).

\bibitem{aknr}
V.A.\ Kosteleck\'y and N.\ Russell,
Phys.\ Lett.\ B {\bf 693}, 443 (2010).

\bibitem{cmm}
D.\ Colladay, P.\ McDonald, and D.\ Mullins,
J.\ Phys.\ A {\bf 43}, 275202 (2010).

\bibitem{jsmv}
J.\ Sk\'akala and M.\ Visser,
Int.\ J.\ Mod.\ Phys.\ D {\bf 19}, 1119 (2010).

\bibitem{nm}
N.\ Mavromatos,
Phys.\ Rev.\ D {\bf 83}, 025018 (2010).

\bibitem{cjlw}
Z.\ Chang, X.\ Li and S.\ Wang,
arXiv:1201.1368;
Z.\ Chang, Y.\ Jiang, and H.\ Lin,
arXiv:1201.3413;
Z.\ Chang and S.\ Wang,
arXiv:1204.2478.

\bibitem{ps}
P.\ Stavrinos,
arXiv:1202.3882.

\bibitem{rgt}
R.G.\ Torrom\'e,
arXiv:1207.3791.

\bibitem{lph}
C.\ L\"ammerzahl, V.\ Perlick, and W.\ Hasse,
arXiv:1208.0619.

\bibitem{kss}
A.P.\ Kouretsis, M.\ Stathakopoulos, and P.C.\ Stavrinos,
arXiv:1208.1673.

\bibitem{qbakpn}
Q.G.\ Bailey and V.A.\ Kosteleck\'y,
Phys.\ Rev.\ D {\bf 74}, 045001 (2006).

\bibitem{lvmm}
R.\ Bluhm \etal,
Phys.\ Rev.\ D {\bf 77}, 065020 (2008).

\bibitem{cardinal}
V.A.\ Kosteleck\'y and R.\ Potting,
Phys.\ Rev.\ D {\bf 79}, 065018 (2009);
Gen.\ Rel.\ Grav.\ {\bf 37}, 1675 (2005).

\bibitem{mds}
M.D.\ Seifert,
Phys.\ Rev.\ Lett.\ {\bf 105}, 0201601 (2010);
Phys.\ Rev.\ D {\bf 82}, 125015 (2010).

\bibitem{deicke}
A.\ Deicke,
Arch.\ Math.\ {\bf 4}, 45 (1953).

\bibitem{ss}
V.S.\ Sabau and H.\ Shimada,
Rep.\ Math.\ Phys.\ {\bf 47}, 31 (2001).

\bibitem{shenpc}
Z.\ Shen, private communication.

\bibitem{sczs}
S.-S.\ Chern and Z.\ Shen,
{\it Riemann-Finsler Geometry},
World Scientific, Singapore, 2005.

\bibitem{js}
M.A.\ Javaloyes and M.\ S\'anchez,
arXiv:1111.5066.

\bibitem{shibata}
C.\ Shibata, 
J.\ Math.\ Kyoto Univ.\ {\bf 24}, 163 (1984).

\bibitem{mh}
M.\ Matsumoto,
Tensor, NS {\bf 24}, 29 (1972);
M.\ Matsumoto and S.\ H\=oj\=o,
Tensor, NS {\bf 32}, 225 (1978).

\bibitem{dcak}
D.\ Colladay and V.A.\ Kosteleck\'y,
Phys.\ Rev.\ D {\bf 55}, 6760 (1997);
Phys.\ Rev.\ D {\bf 58}, 116002 (1998).

\bibitem{tables}
{\it Data Tables for Lorentz and CPT Violation},
V.A.\ Kosteleck\'y and N.\ Russell,
Rev.\ Mod.\ Phys.\ {\bf 83}, 11 (2011)
[arXiv:0801.0287].

\bibitem{rl-redef}
R.\ Lehnert,
Phys.\ Rev.\ D {\bf 74}, 125001 (2006);
Rev.\ Mex.\ Fis.\ {\bf 56}, 469 (2010).

\bibitem{redef}
D.\ Colladay and P.\ McDonald,
J.\ Math.\ Phys.\ {\bf 43}, 3554 (2002);
M.S.\ Berger and V.A.\ Kosteleck\'y,
Phys.\ Rev.\ D {\bf 65}, 091701(R) (2002);
V.A.\ Kosteleck\'y and M.\ Mewes,
Phys.\ Rev.\ D {\bf 66}, 056005 (2002);
Q.G.\ Bailey and V.A.\ Kosteleck\'y,
Phys.\ Rev.\ D {\bf 70}, 076006 (2004);
B.\ Altschul,
J.\ Phys.\ A {\bf 39} 13757 (2006);
V.A.\ Kosteleck\'y and J.D.\ Tasson,
Phys.\ Rev.\ Lett.\ {\bf 102}, 010402 (2009);
Phys.\ Rev.\ D {\bf 83}, 016013 (2011).

\bibitem{km-nonmin}
V.A.\ Kosteleck\'y and M.\ Mewes,
Ap.\ J.\ Lett.\ {\bf 689}, L1 (2008);
Phys.\ Rev.\ D {\bf 80}, 015020 (2009);
Phys.\ Rev.\ D {\bf 85}, 261603 (2012).

\bibitem{rl-fermion}
R.\ Lehnert,
J.\ Math.\ Phys.\ {\bf 45}, 2299 (2004).

\bibitem{badc}
B.\ Altschul and D.\ Colladay,
Phys.\ Rev.\  D {\bf 71}, 125015 (2005).

\bibitem{gls}
F.\ Girelli, S.\ Liberati, and L.\ Sindoni,
Phys.\ Rev.\ D {\bf 75}, 064015 (2007).

\bibitem{abk}
B.\ Altschul \etal,
Phys.\ Rev.\ D {\bf 81}, 065028 (2010).

\bibitem{akrl}
V.A.\ Kosteleck\'y and R.\ Lehnert,
Phys.\ Rev.\ D {\bf 63}, 065008 (2001).

\bibitem{electron}
B.R.\ Heckel \etal,
Phys.\ Rev.\ D {\bf 78}, 092006 (2008);
R.\ Bluhm and V.A.\ Kosteleck\'y,
Phys.\ Rev.\ Lett.\ {\bf 84}, 1381 (2000);
W.A.\ Terrano, B.R.\ Heckel, and E.G.\ Adelberger,
Class.\ Quant.\ Grav.\ {\bf 28}, 145011 (2011);
Y.\ Bonder and D.\ Sudarsky,
Class.\ Quant.\ Grav.\ {\bf 25}, 105017 (2008).

\bibitem{neutron}
F.\ Can\`e \etal,
Phys.\ Rev.\ Lett.\ {\bf 93}, 230801 (2004);
B.\ Altschul,
Phys.\ Rev.\ D {\bf 79}, 061702 (R) (2009).

\bibitem{muon}
G.W.\ Bennett \etal,
Muon $g$--2 Collaboration,
Phys.\ Rev.\ Lett.\ {\bf 100}, 091602 (2008);
R.\ Bluhm \etal,
Phys.\ Rev.\ Lett.\ {\bf 84}, 1098 (2000).

\bibitem{neutrino}
V.A.\ Kosteleck\'y and M.\ Mewes,
Phys.\ Rev.\ D {\bf 85}, 096005 (2012);
J.S.\ D\'\i az \etal,
Phys.\ Rev.\ D {\bf 80}, 076007 (2009).

\bibitem{zs}
Z.\ Shen,
Canad.\ J.\ Math.\ {\bf 55}, 112 (2003).

\bibitem{dbcr}
D.\ Bao and C.\ Robles,
in D.\ Bao, R.L.\ Bryant, S.-S.\ Chern, and Z.\ Shen, eds.,
{\it A Sampler of Riemann-Finsler Geometry},
Cambridge University Press, Cambridge, 2004.

\end{thebibliography}
\end{document}